\documentclass{aa}  

\usepackage{graphicx}
\usepackage{txfonts}
\usepackage{color}
\usepackage[]{hyperref}
\hypersetup{
    colorlinks=true,
    linkcolor=blue,
    filecolor=blue,      
    urlcolor=blue,
    citecolor=blue
}
\usepackage{multicol}
\usepackage{color}
\usepackage{braket}
\usepackage{autobreak}
\usepackage{enumitem}
\usepackage{mathrsfs}
\usepackage{amsfonts}
\usepackage{amstext}
\usepackage{amsmath}
\usepackage{xcolor} 
\usepackage{natbib}
\usepackage{subfigure}
\def\fl#1 {\textcolor{cyan}{#1}\;}
\newcommand{\hi } {{\rm H}\,{\small\rm I}}

\begin{document}

   \title{The baryonic mass-size relation of galaxies. II. Implications for the evolutionary paths between star-forming and passive galaxies}
   \titlerunning{The baryonic mass-size relation of galaxies. II.}

\author{Zichen Hua \inst{1, 2, 3},
          Federico Lelli \inst{1},
          Enrico Di Teodoro\inst{4, 1},
          Stacy McGaugh\inst{5},
          James Schombert\inst{6}
          }

   \institute{Arcetri Astrophysical Observatory, INAF, Largo Enrico Fermi 5, 50125, Florence, Italy
        \and
   Department of Astronomy, University of Science and Technology of China, Hefei 230026, China
        \and
            School of Astronomy and Space Sciences, University of Science and Technology of China, Hefei 230026, China
        \and
             Dipartimento di Fisica e Astronomia, Universit$\rm \grave{a}$ degli Studi di Firenze, 50019 Sesto Fiorentino, Italy
        \and
            Department of Astronomy, Case Western Reserve University, 10900 Euclid Avenue, Cleveland, OH 44106, USA
        \and
            Department of Physics, University of Oregon, Eugene, OR 97403, USA
        }
   \authorrunning{Z. Hua et al.}
   \date{Received 20 October 2025/ Accepted 15 January 2026}

  \abstract{
  The baryonic mass-size relation of galaxies links the total baryonic mass (stars plus gas) to the baryonic half-mass radius. In the first paper of this series, we showed that star-forming galaxies from the SPARC sample follow two distinct relations in the baryonic mass-size plane: one defined by high-surface-density (HSD), star-dominated, Sa-to-Sc galaxies, and one defined by low-surface-density (LSD), gas-dominated, Sd-to-dI galaxies. In this second paper, we study the structural relations between baryonic mass, half-mass radius, and mean surface density to constrain possible morphological transformations between star-forming and passive galaxies. We complemented the SPARC sample with $\sim$1200 passive galaxies that are nearly devoid of gas: ellipticals (Es), lenticulars (S0s), dwarf ellipticals (dEs) or dwarf spheroidals (dSphs), and the so-called `ultra-diffuse galaxies' (UDGs). Our results can be summarised as follows: (1) \textcolor{black}{passive stellar components follow four distinct relations at high statistical significance, namely (i) ellipticals plus bulges, (ii) S0 disks, (iii) non-nucleated dwarfs (dEs, dSphs, UDGs), and (iv) nucleated dEs;} (2) star-forming HSD disks (mostly Sa to Sc) overlap with S0 disks \textcolor{black}{within 2$\sigma$ in the baryonic relations and within 1$\sigma$ in the stellar ones}, so \textcolor{black}{present-day} spirals may simply evolve into lenticulars as they run out of gas; (3) star-forming LSD disks (mostly Sd to dI) \textcolor{black}{are offset from non-nucleated passive dwarfs at more than 3$\sigma$ in the baryonic relations}, but the two galaxy populations overlap \textcolor{black}{within 1$\sigma$} in the stellar \textcolor{black}{relations}, suggesting that \textcolor{black}{non-nucleated} passive dwarfs may form from star-forming dwarfs \textcolor{black}{only} after gas removal; (4) UDGs \textcolor{black}{extend} the sequence of non-nucleated dEs/dSphs and may originate from \textcolor{black}{the most diffuse} star-forming LSD galaxies with no need for a substantial expansion of the stellar component.}


   \maketitle

\section{Introduction}\label{sec_intro}

A classic approach to investigate the formation and evolution of galaxies is studying the structural relationships between their total luminosity (or total stellar mass, $M_\star$), half-light radius (or half-mass radius, $R_{50, \star}$), and mean surface brightness (or mean stellar surface density, $\Sigma_\star = M_\star/2\pi R_{50, \star}^2$). For example, passive `quenched' galaxies in galaxy clusters follow different luminosity$-$size$-$surface brightness relationships \citep[e.g.][]{Kormendy1985-dE, Binggeli1991, Bingelli1994}: one set by giant ellipticals (Es) and one by dwarf ellipticals (dEs). This fact indicates that dEs are not a `miniature' version of giant Es, despite their similar morphology and resulting historical nomenclature. Rather, dEs follow the same structural relations of dimmer dwarf spheroidals (dSphs) in the Local Group and in other nearby groups. This evidence led \citet{Kormendy2009-Sph} to rename passive dwarfs (dEs or dSphs) as `spheroidals' in contrast to proper `ellipticals'. In this paper, we use the historical nomenclature dEs/dSphs, but there is no clear-cut distinction between dEs and dSphs. In addition, \citet{Kormendy2012} proposed a revised Hubble tuning fork \citep[following early ideas of][]{vandenBergh1976-classification} in which passive lenticulars (S0s) with different bulge-to-disk ratios (S0a, S0b, S0c) form a parallel sequence to the classic one of star-forming spirals (Sa, Sb, Sc). At the end of this revised Hubble sequence, there are `bulgeless' low-mass galaxies: dSphs/dEs on the passive branch and dwarf irregulars (dIs) on the star-forming branch.

In the first paper of this series  \citep[][hereafter Paper I]{Hua2025-MR}, we introduced the baryonic mass-size relation of galaxies, which links the total baryonic mass ($M_{\rm bar} = M_\star + M_{\rm gas}$) to the baryonic half-mass radius ($R_{\rm 50, bar}$), enclosing half of $M_{\rm bar}$. In passive galaxies (Es, S0s, dEs/dSphs), the gas mass ($M_{\rm gas}$) is often negligible (but see \citealt{Serra2012}), so the baryonic mass-size plane is equivalent to the `classic' stellar mass-size plane. The situation is very different for star-forming galaxies, especially for dwarf galaxies in which $M_{\rm gas}$ can be much higher than $M_\star$ \citep[e.g.][]{Lelli2022}. In Paper\,I, we found that star-forming galaxies follow two distinct sequences in the $M_{\rm bar}-R_{\rm 50, bar}$ plane: one defined by star-dominated, high-surface-density (HSD) galaxies (mostly Sa to Sc) and one by gas-dominated, low-surface-density (LSD) galaxies (mostly Sd to dI). These two sequences were previously found in the luminosity-size plane by \citet{Schombert2006-structure} and are reminiscent of the dichotomy between \textcolor{black}{high-mass passive galaxies (Es and S0s) and low-mass ones (dEs and dSphs)}.

In this second paper, we constrain the possible morphological transformations between star-forming and passive galaxies by studying the structural relations between the baryonic quantities $M_{\rm bar}$, $R_{\rm 50, bar}$, and $\Sigma_{\rm 50, bar}= M_{\rm 50, bar}/2\pi R_{\rm 50, bar}^2$, as well as the stellar quantities $M_{\rm \star}$, $R_{\rm 50, \star}$, and $\Sigma_{\rm 50, \star}$. The basic idea is simple. If a galaxy converts its gas into stars in a close-box fashion, it will move across the stellar relations but stay nearly at the same location of the baryonic ones (apart from minor variations in $R_{\rm 50, bar}$ depending on where the star formation occurs). Instead, if a galaxy suddenly loses its gas, it will move across the baryonic relations but stay almost at the same location of the stellar ones, unless the gas loss occurs together with a significant change in the stellar distribution. This simple idea provides key insights in galaxy evolution, especially in the dwarf regime where the gas content can be either totally dominant or totally negligible.

This paper is structured as follows. In Section~\ref{sec_method}, we describe our galaxy sample and data analysis. In Section~\ref{sec_results}, we present the baryonic structural relations for passive and star-forming galaxies. In Section~\ref{sec_diss}, we discuss our results in terms of galaxy formation and evolution. In Section~\ref{sec_conclusion}, we provide a brief summary.

\section{Datasets}\label{sec_method}

For star-forming galaxies, we use the same sample as in Paper\,I, which consists of 169 galaxies (Sa to dI) from the SPARC database \citep{Lelli2016-SPARC}. For the derivation of $M_{\rm bar}$ and $R_{\rm 50, bar}$, we refer to Paper I. We just recall that stellar masses are measured using \textit{Spitzer} photometry at 3.6 $\mu$m and assuming a mass-to-light ratio ($\Upsilon_{\star}^{[3.6]}$) of 0.5 for the stellar disk and 0.7 for the stellar bulge. \textcolor{black}{These values are derived using the stellar population synthesis (SPS) models by \citet{Schombert2014-SPS} and subsequent improvements by \citet{Schombert2019-SFH}, assuming the \citet{Chabrier2003-IMF} initial mass function (IMF) and the chemical-enrichment model of \citet{Prantzos2009-chemical}.} Variations in $\Upsilon_\star$ from galaxy to galaxy are taken into account in the error budget, assuming a scatter of 25$\%$. \textcolor{black}{Recent studies \citep[e.g.][]{Schombert2022} show that $\Upsilon_\star^{[3.6]}$ of LSD disks might be smaller ($\sim$0.4 on average) than that of HSD disks ($\sim$0.5 on average) due to differences in their star-formation and chemical enrichment histories (see Fig. 1 of \citealt{Schombert2022}). However, such a small difference in $\Upsilon_{\star}^{[3.6]}$ has negligible effects to our final results because the variation in $M_\star$ is smaller than the uncertainties.}

In the following, we describe the data collection for passive galaxies with no substantial star-formation activity (Es, S0s, dEs/dSphs, UDGs). Even if these galaxies are not forming stars at a substantial rate, some of them may contain atomic and/or molecular gas \citep[e.g.][]{Serra2012, Shelest2020}. However, the gas fraction is generally very small ($M_{\rm gas}/M_{\rm bar} \lesssim 0.01$), \textcolor{black}{so we neglect any gas contribution and assume that $M_{\rm bar}\simeq M_{\rm \star}$ and $R_{\rm 50, bar}\simeq R_{\rm 50, \star}$ to a first-order approximation.}

\subsection{Ellipticals and lenticulars}

We considered 140 ellipticals from \citet{Schombert2016-Spitzer_E} that have \textit{Spitzer} images at $3.6 \ {\rm \mu m}$. This dataset is fully consistent with the SPARC dataset in terms of photometric band and photometric procedures. To derive $M_{\rm \star}$, we adopted $\Upsilon_{\star}^{[3.6]} = 0.9$ M$_\odot$/L$_\odot$ \citep{Lelli2017-RAR, Schombert2022}. This value comes from the same SPS models that set $M_{\rm \star}$ for SPARC galaxies, so the two datasets are on a common stellar mass scale. \textcolor{black}{The difference in $\Upsilon_{\star}^{[3.6]}$ between elliptical and star-forming galaxies is largely driven by differences in their star-formation and chemical-enrichment histories \citep[e.g.][]{Schombert2022}}.

In addition, we built a sample of 33 lenticulars (S0s) that have available bulge-disk decompositions from the literature:
\begin{enumerate}

\item sixteen S0s from \citet{Lelli2017-RAR} with \textit{Spitzer} images at 3.6 $\mu$m. To derive $M_{\rm \star}$ we adopted $\Upsilon_{\rm bul}^{[3.6]} = \Upsilon_{\rm disk}^{[3.6]} = 0.8$ M$_\odot$/L$_\odot$ \citep{Lelli2017-RAR}. \textcolor{black}{This value comes from the same SPS models that set $M_{\rm \star}$ for SPARC galaxies, so the datasets are} on the same stellar-mass scale.

\item ten S0s from \citet{Rizzo2018-S0}. This work provides three different estimates of $M_{\rm \star}$. For consistency with the SPARC sample, we used $M_\star$ from the $K$-band luminosity assuming $H_0 = 73$ km s$^{-1}$ Mpc$^{-1}$ and $\Upsilon^{\rm K}_{\star}=1$ for both bulge and disk. This value is consistent with the SPARC stellar-mass scale because $\Upsilon^{\rm K}_{\star}\simeq1.29 \Upsilon^{[3.6]}_\star$ with no significant colour term \citep{McGaugh2014}, so $\Upsilon^{[3.6]}_\star=0.8$ translates into $\Upsilon^{\rm K}_\star\simeq1.0$. The uncertainties in $M_\star$ from K-band data are dominated by galaxy-to-galaxy variations in $\Upsilon_\star^{\rm K}$ and are comparable to those from [3.6] data. To derive $M_{\rm bul}$ and $M_{\rm disk}$ separately, we used the $r-$band $B/T$ from \citet{Rizzo2018-S0}. $R_{50, \rm bul}$ and $R_{50, \rm disk}$ were estimated by a parametric decomposition in the $r-$band as described by \citet{Rizzo2018-S0}.

\item seven S0s from \citet{Kormendy2012}. Specifically, we select only S0s with `classical bulges' to be consistent with the SPARC non-parametric bulge-disk decompositions, which assigned `pseudo-bulges' to the disk component \citep[see][]{Lelli2016-SPARC}. \citet{Kormendy2012} provide the $V-$band absolute magnitude, so we compute $M_{\rm bul}$ and $M_{\rm disk}$ assuming $\Upsilon_{\rm bul}^{\rm V} = \Upsilon_{\rm disk}^{\rm V}= 5$ M$_\odot/L_\odot$ in the $V-$band \citep{Schombert2022}. This value comes from the same SPS models that set $M_\star$ in SPARC galaxies, so the galaxies are on a common stellar-mass scale, but the uncertainties in $M_\star$ from $V$-band data are significantly larger than those from [3.6] or $K$-band data.

\end{enumerate}

\subsection{Dwarf ellipticals and dwarf spheroidals}

We considered passive dwarfs (dEs/dSphs) in the core regions of galaxy clusters, so we can assume that they are largely devoid of gas. In particular, we consider galaxies with available photometry from the following surveys:
\begin{enumerate}
    \item 404 galaxies from the Next Generation Virgo Cluster Survey \citep[NGVS,][]{Ferrarese2019-NGVS}. These galaxies are identified as members of the Virgo cluster by visual inspections \citep{Ferrarese2019-NGVS}, so they are the most reliable.
    $R_{\rm 50, \star}$ was derived from the curve of growth based on the $g-$band surface brightness profiles.
    \item 627 galaxies from the Next Generation Fornax Cluster Survey (NGFS), specifically 243 from \citet{Eigenthaler2018-NGFS} and 384 from \citet{OrdenesBriceno2018-NGFS}. $R_{\rm 50, \star}$ was estimated from $i-$band images.
\end{enumerate}

Given the availability of $i-$ and $g-$ magnitudes, we calculated stellar masses using the relation from \citet{Taylor2011-Mstar}:
\begin{equation}\label{eq_MsTaylor}
    \log \left(\frac{M_\star/M_{\rm \odot}}{L_{i}/L_{\rm \odot}}\right) = -0.68 + 0.70 \left(g - i\right).
\end{equation}
\textcolor{black}{Equation (\ref{eq_MsTaylor}) is based on the \citet{Chabrier2003-IMF} IMF and the SPS models of \citet{Bruzual2003-SPS}. For low-mass galaxies with typical colours, Eq. (\ref{eq_MsTaylor}) is virtually the same as the one from \citet{Schombert2014-SPS} \citep[see also Fig. 3 of][]{Schombert2022}.} Therefore, the stellar masses of these passive dwarfs are on the same mass scale as SPARC galaxies. The catalogue from \citet{OrdenesBriceno2018-NGFS} does not provide $(g-i)$ values, so \textcolor{black}{for these 384 dwarfs} we assumed $(g-i) = 0.85$, which is the average \textcolor{black}{color} of dEs/dSphs from \citet{Eigenthaler2018-NGFS}.

We excluded 6 galaxies from the NGVS \citep{Ferrarese2019-NGVS} because they have $M_{\rm \star} > 10^{10} \ M_{\rm \odot}$ and are morphologically classified as Es or S0s. \textcolor{black}{The stellar masses of these galaxies would be even larger using the SPS models of \citet{Schombert2014-SPS}, which give higher $\Upsilon_\star$ than eq.\,(\ref{eq_MsTaylor}) for the red colours typical of Es and S0s.} Thus, we are confident that these objects are not dwarf galaxies. We did not consider them in our sample of Es and S0s for the sake of internal consistency: all the ellipticals in our sample have \textit{Spitzer} data, while all the lenticulars have accurate bulge-disk decompositions.

Finally, we added 9 galaxies from \citet{Kormendy2009-Sph} and one from \citet{Kormendy2012}, which are not included in the NGVS and NGFS catalogues. These galaxies are relatively massive dEs with $M_\star \simeq 10^9$ M$_\odot$ and occupy the interesting region connecting the sequences of dEs and Es. We computed their stellar mass assuming $\Upsilon_{\star}^{\rm V} = 2$ M$_\odot/L_\odot$ in the $V-$band. This value is motivated by studies of resolved stellar populations in nearby dEs/dSphs and puts them on the same stellar-mass scale as SPARC galaxies \citep{Lelli2017-RAR}. Indeed, these 9 dEs/dSphs with $M_\star$ from $V$-band \textcolor{black}{images} do not show any systematics with respect to those with $M_\star$ from $g$-band \textcolor{black}{images}. In conclusion, our final sample of dEs/dSphs contains 1034 galaxies.

\subsection{Ultra-diffuse galaxies}

The term `ultra-diffuse galaxies' (UDGs) has been introduced by \citet{VandeK2015-UDG} to describe galaxies with central surface brightness lower than 24.5 mag arcsec$^{-2}$ in the $g-$band and effective radius larger than 1.5 kpc. It is debated whether UDGs are a new type of galaxies (such as `failed $L_\star$ galaxies', \citealt{VandeK2015-UDG}), or merely the extension of usual dwarf galaxies to lower surface brightnesses \citep[e.g.][]{Conselice2018, Chilingarian2019, Iodice2020, Marleau2021, Rong2017-UDG, Rong2024-UDG, Li2023-UPG, Wright2021-UDG, Zoller2024, Lelli2024-UDG}. In the following, we use the term `UDGs' to simply distinguish them from the historical population of dEs/dSphs. 

We considered 46 UDGs in the Coma cluster \citep{VandeK2015-UDG} and \textcolor{black}{21} UDGs in other clusters, as compiled by \citet{Gannon2024-UDG}. These cluster UDGs can be assumed to be devoid of gas, so that $M_{\rm bar}\simeq M_{\rm \star}$ and $R_{\rm 50, bar}\simeq R_{\rm 50, \star}$. For the UDGs in Coma, $R_{\rm 50,\star}$ was measured from $g-$band images and $M_{\rm \star}$ was derived using Eq.~(\ref{eq_MsTaylor}) with $g-i = 0.8$ \citep{VandeK2015-UDG}, similarly to usual dEs/dSphs. For the UDGs from \citet{Gannon2024-UDG}, both $R_{\rm 50, \star}$ and $M_{\rm \star}$ were taken directly from their catalogue. The data from \citet{Gannon2024-UDG} are approximately on the same mass scale as the rest of our sample, showing no strong systematics.

\begin{figure*}
    \centering
    \includegraphics[width=\linewidth]{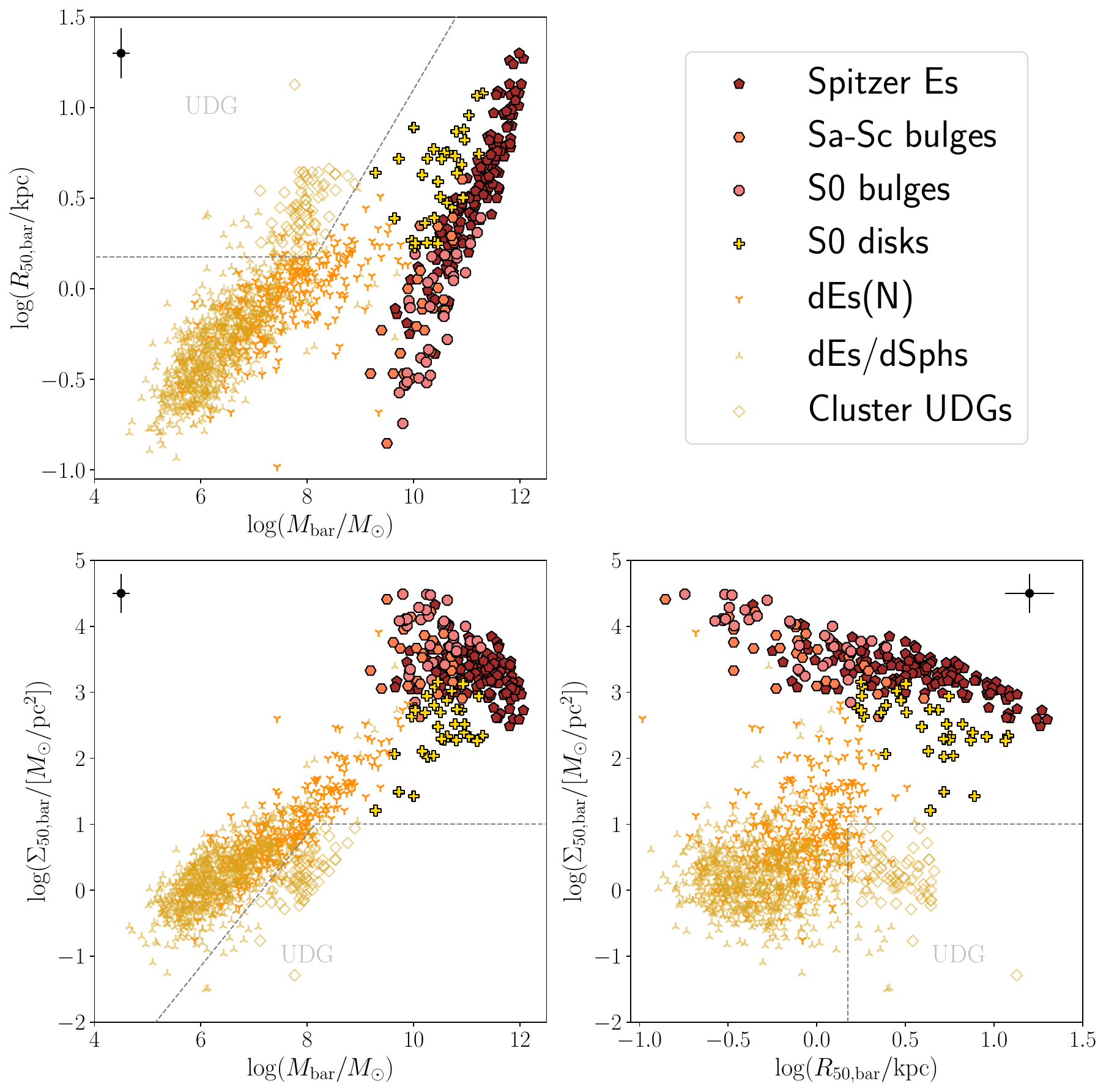}
    \caption{Baryonic structural relations for passive galaxies: $M_{\rm bar}$ vs $R_{\rm 50, bar}$ (top left), $M_{\rm bar}$ vs $\Sigma_{\rm 50, bar}$ (bottom left), and $R_{\rm 50, bar}$ vs $\Sigma_{\rm bar}$ vs (bottom right). See Sect.\,\ref{sec_method} for references to the various datasets of Es, S0s, dEs/dSphs, and UDGs. We also considered the `classical' bulges of Sa-Sc galaxies. \textcolor{black}{The mean uncertainties of our sample galaxies are indicated in the top corner of each panel.} For all these objects, the baryonic structural relations are virtually the same as the stellar ones. The grey-dashed lines show the selection criteria of UDGs from \protect{\citet{VandeK2015-UDG}}.}
    \label{fig1}
\end{figure*}

\begin{figure}
    \centering
    \includegraphics[width=\linewidth]{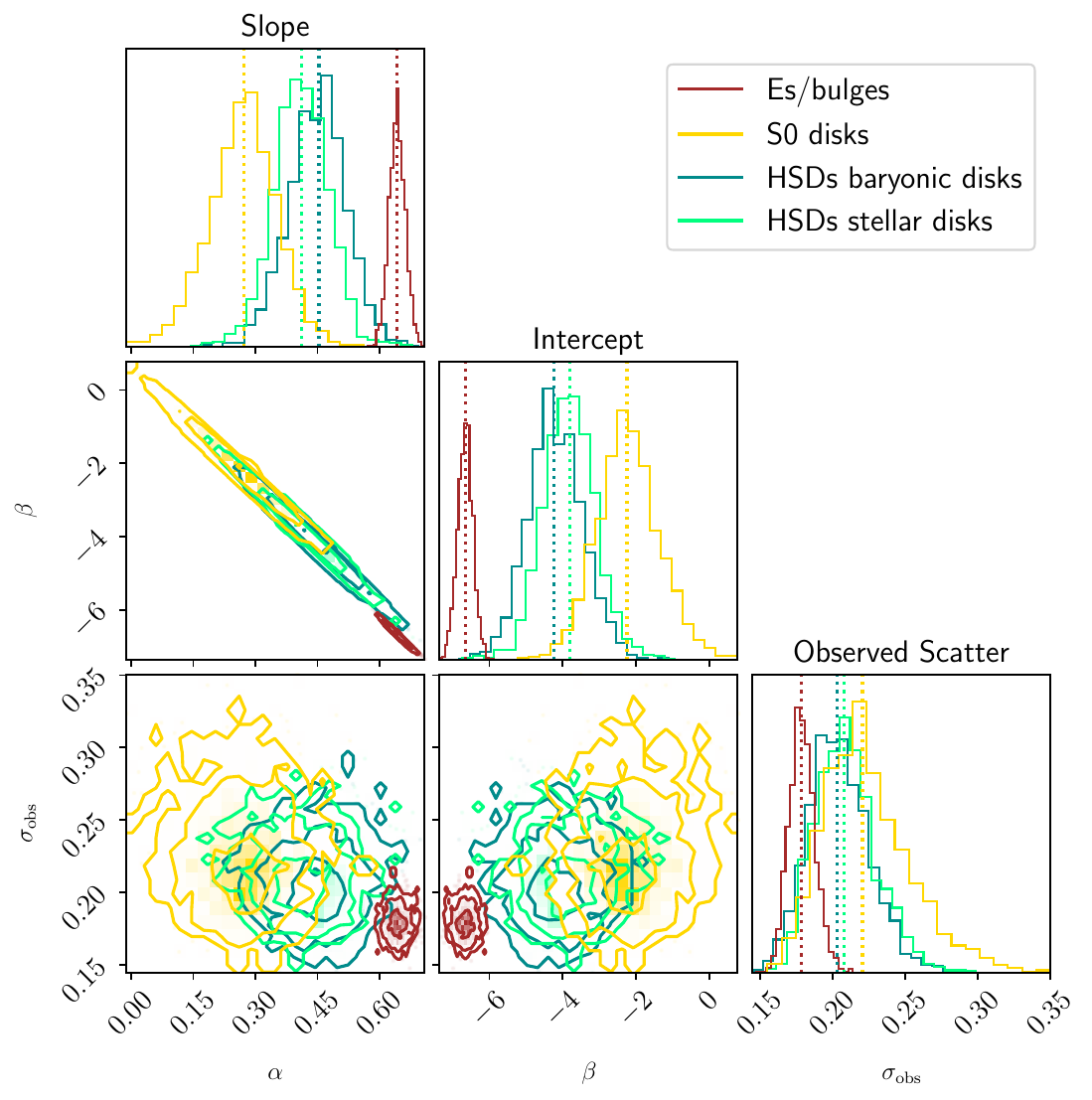}
    \caption{\textcolor{black}{Corner plots of the mass-size relations of Es and bulges, S0 disks, HSD baryonic disks, and HSD stellar disks. The panels show the 2D posterior probability distributions of pairs of fitting parameters, and the 1D marginalised probability distribution of each fitting parameter (histograms). In the 2D distributions, the contours correspond to the $1\sigma$, $2\sigma$ and $3\sigma$ regions, respectively. In the histograms, the dashed lines correspond to the median values.}}
    \label{fig_corner_HSD}
\end{figure}

\begin{figure}
    \centering
    \includegraphics[width=\linewidth]{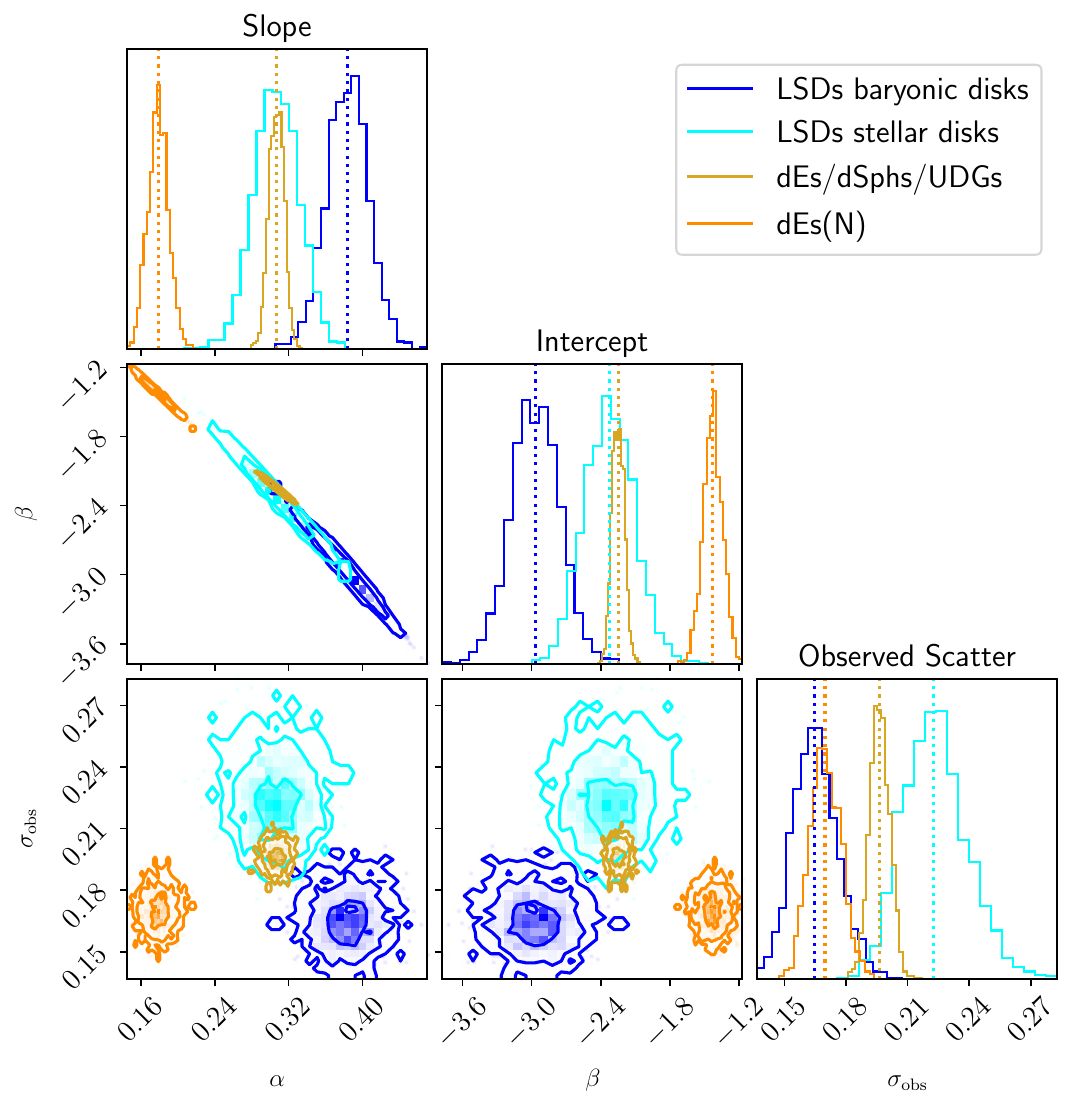}
    \caption{\textcolor{black}{Same as Fig.~\protect{\ref{fig_corner_HSD}} but for non-nucleated dEs/dSphs/UDGs, nucleated dEs(N), baryonic LSD disks, and stellar LSD disks.}}
    \label{fig_corner_LSD}
\end{figure}

\begin{figure*}
    \centering
    \includegraphics[width=\linewidth]{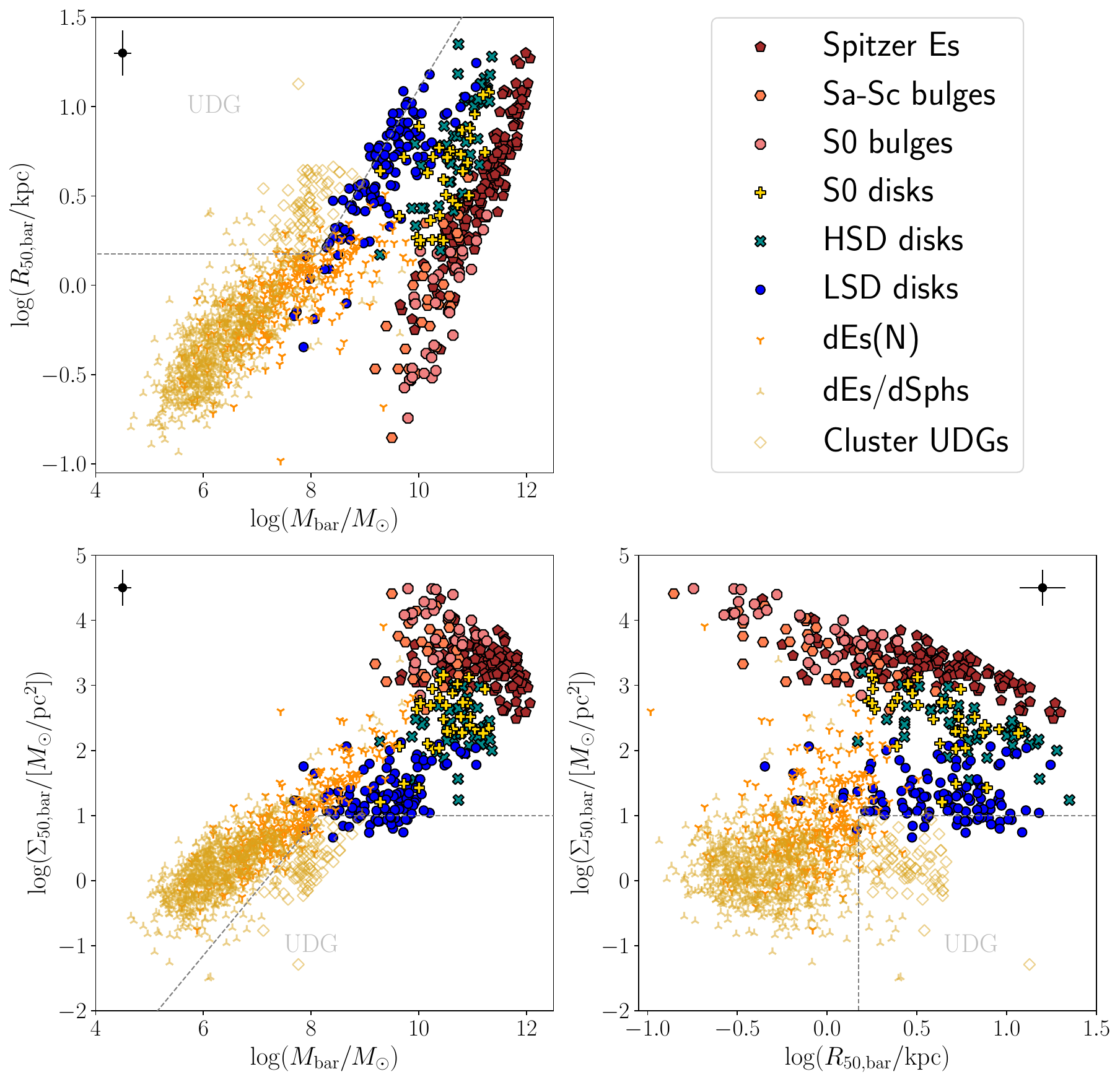}
    \caption{Same as Fig.\,\ref{fig1} but adding star-forming disks from the SPARC database. As we show in Paper I, star-forming disks define two distinct sequences: one formed by star-dominated HSD galaxies (mostly Sa-Sc) and one by gas-dominated LSD galaxies (mostly Sd-dI).}
    \label{fig2}
\end{figure*}

\begin{figure*}
    \centering
    \includegraphics[width=\linewidth]{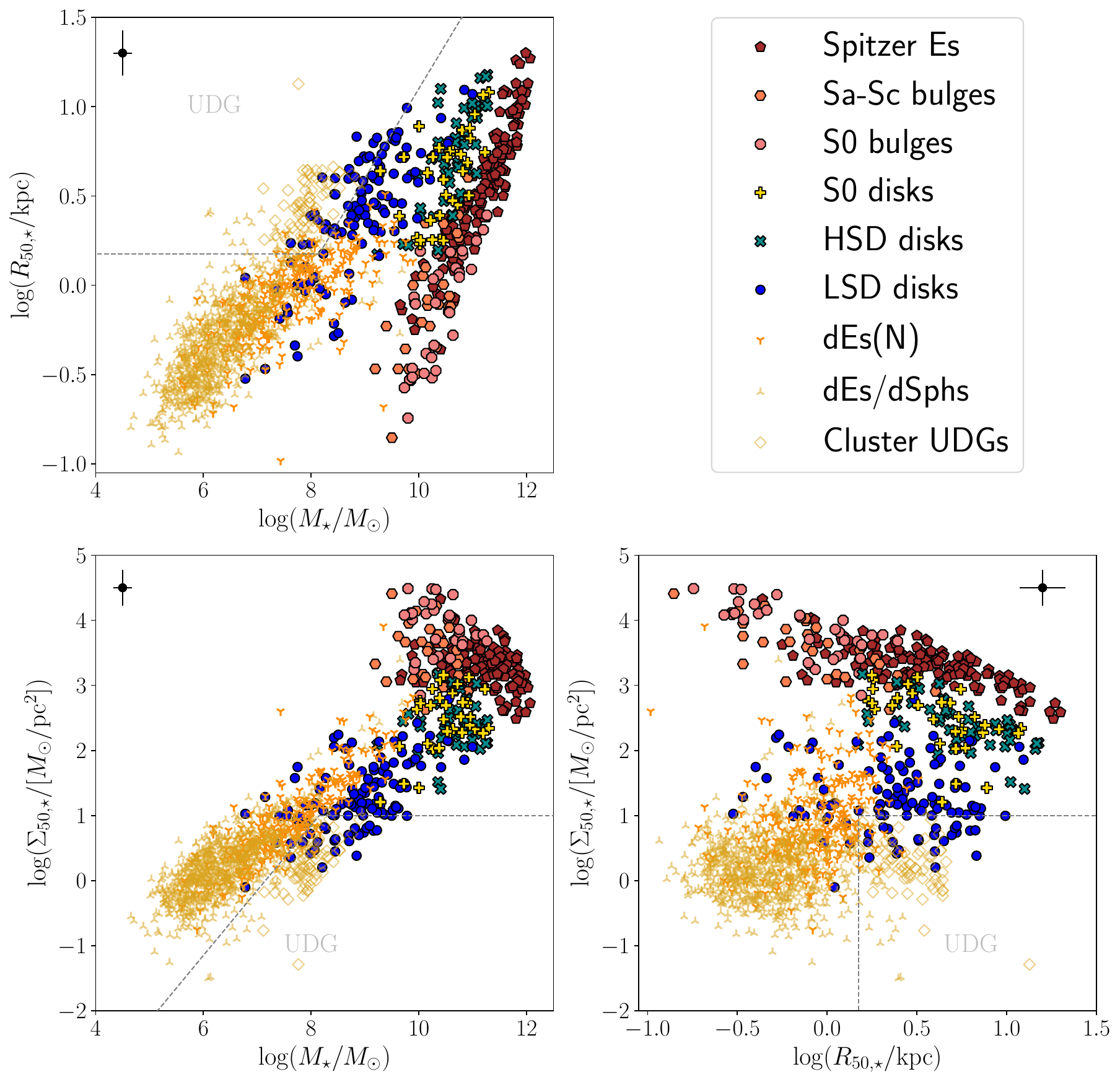}
    \caption{Same as Fig.\,\ref{fig2} but considering stellar quantities rather than baryonic ones. The main difference occurs for gas-dominated LSD disks, which move closer to the sequence defined by passive dwarfs (dEs, dSphs, UDGs).}
    \label{fig3}
\end{figure*}

\section{Results}\label{sec_results}

In this section, we present the `structural planes' between baryonic mass $M_{\rm bar}$, half-mass radius $R_{\rm 50, bar}$, and mean surface density within the half-mass radius $\Sigma_{\rm 50, bar}$. Firstly, we discuss passive galaxies for which the baryonic and stellar quantities are virtually the same \textcolor{black}{(Fig.\,\ref{fig1})}, so we largely revisit known observational evidence. Subsequently, we discuss star-forming galaxies considering both their baryonic and stellar quantities \textcolor{black}{(Fig.\,\ref{fig2} and Fig.\,\ref{fig3}, respectively)}.

\textcolor{black}{The various structural relations followed by different galaxy populations were fitted using the \textsc{roxy} package \citep{Bartlett2023-roxy}, which performs a Bayesian linear fit using a Markov-Chain Monte Carlo (MCMC) technique (see Appendix\,\ref{app_1} for technical details). The fit results are summarised in Table~\ref{tab_star_scaling}. Posterior probability distributions of the fitting parameters are shown for the mass-size relations (Fig.\,\ref{fig_corner_HSD} and Fig.\,\ref{fig_corner_LSD}).}

\subsection{Passive galaxies}

\subsubsection{\textcolor{black}{Confirming the E$-$dE dichotomy}}\label{sec_MR_highmass}

Figure\,\ref{fig1} shows the structural relations of passive galaxies (Es, S0s, dEs/dSphs, UDGs). The bulges and disks of S0s are considered to be separate components. We also add the bulges of spiral galaxies, which may be considered as passive spheroids. In particular, we recall that the bulge-disk decompositions from the SPARC dataset \citep{Lelli2016-SPARC} consider only `classical' bulges (mostly found in Sa-Sb) and assign other structures (bars, barlenses, pseudobulges) to the stellar disk because of their common origin and similar $\Upsilon_\star$. 

The dichotomy between Es and dEs/dSphs \citep{Kormendy1985-dE, Binggeli1991} is evident. In the stellar or baryonic planes, the dichotomy is more emphasised than in the photometric planes \citep[cf. with][]{Kormendy2009-Sph} because Es tend to have higher $\Upsilon_\star$ than dEs/dSphs. Stellar bulges form a continuous sequence with Es but are distinct from dEs/dSphs, in agreement with previous studies \citep{Kormendy2009-Sph, Kormendy2012}. There exists a rarer population of `bright' dEs that connects the two sequences at $\log(M_\star/M_\odot)\simeq9-10$ and $\Sigma_{\rm 50, bar}\simeq 100-100$ M$_\odot$ pc$^{-2}$  \citep[see also][]{Graham2003, Gavazzi2005}. The key point of the E-dE dichotomy, however, is not the presence (or absence) of a gap in the structural planes, but the fact that Es/bulges and dEs/dSphs follow relationships with different slopes. Indeed, we find that Es and bulges share similar slopes of $\sim$0.6, while dEs/dSphs have slopes $\sim0.2-0.3$ \textcolor{black}{(see Table~\ref{tab_star_scaling}, Fig.\,\ref{fig_corner_HSD} and Fig.\,\ref{fig_corner_LSD})}. Some authors proposed that passive galaxies form a continuous but non-monotonic sequence in the stellar mass-size plane \citep[e.g.][]{Eigenthaler2018-NGFS}; the change in slope must then correspond to different physical regimes where different processes are in action.

There are three dEs/dSphs from NGVS \citep{Ferrarese2019-NGVS} \textcolor{black}{towards the bottom of} the Es$+$bulge sequence. These three galaxies were classified as `ellipticals' by \citet{Kormendy2009-Sph}; we consider them in the dEs/dSphs sample just to keep them distinct from the \textit{Spitzer} Es from \citet{Schombert2016-Spitzer_E}. Based on their structural properties, these three galaxies appear to be similar to `compact ellipticals' like M32. The astronomical jargon here may become very confusing: `compact ellipticals' extend the elliptical sequence to low masses, so one may be tempted to call them `dwarf ellipticals', but the name dEs has historically been used to refer to low-mass galaxies on the other sequence \citep[the `dwarf sequence', following][]{Bingelli1994}. Arguably, the nomenclature `Es' versus `Sphs' advocated by \citet{Kormendy2009-Sph} is simpler and more appropriate, but we adhere to the historical nomenclature in this paper.

\subsubsection{\textcolor{black}{Nucleated and non-nucleated dwarfs}}\label{sec_nucleated}

Figure\,\ref{fig1} distinguishes between nucleated dwarfs, indicated as dEs(N), and non-nucleated dwarfs, which are hereafter simply indicated as dEs/dSphs. Nucleated dwarfs tend to have systematically higher masses and mean surface densities than non-nucleated ones. If we fit the mass-size relation of dEs/dSphs and dEs(N) separately, the former one has a significantly steeper slope ($ 0.26\pm 0.01$) than the latter one ($0.20 \pm 0.01$). \textcolor{black}{The posterior probability distributions of the fitting parameters (Fig.~\ref{fig_corner_LSD}) confirm that the mass-size relations of these two populations are inconsistent with each other at more than $3\sigma$.} As suggested by \citet{Jerjen1997}, nucleated dEs(N) may form a different `breed' than non-nucleated dEs, which are instead of the same `breed' as dSphs and UDGs. In particular, dEs(N) generally have Sersic indexes ($n_{\rm s}$) higher than 1 \citep{Binggeli1991}, which progressively increase with stellar mass or luminosity, forming a single monotonic $n_{\rm s}-M_\star$ relation together with Es \citep{Jerjen1997}. 

The mass-size relation of UDGs naturally extends that of dEs/dSphs. This fact, which is evident by eye, is further confirmed by their \textcolor{black}{roughly} consistent best-fit slopes: $0.18 \pm 0.06$ for UDGs and $0.26 \pm 0.01$ for dEs/dSphs (see Table~\ref{tab_star_scaling}). We notice that a few dEs/dSphs meet the selection criteria of UDGs (see the grey-dashed lines in Fig.~\ref{fig1}, \citealp{VandeK2015-UDG}). Some of them, indeed, are considered as UDGs in a few studies \citep[e.g.][]{Lim2020-UDG, Venhola2017-UDG}. However, even if we remove these ambiguous galaxies, the best-fit slope of dEs/dSphs is still consistent with that of UDGs. This suggests that UDGs do not constitute a distinct population from passive (non-nucleated) dwarfs, as it has already been pointed out by several authors \citep{Conselice2018, Chilingarian2019, Marleau2021, Zoller2024}. Notably, current surveys seem to lack and/or under-represent more extreme dwarfs with $\log(M_{\rm bar}/M_\odot)\simeq10^5-10^7$, $R_{\rm 50, bar}>1$ kpc, and $\Sigma_{\rm 50, bar} < 1$ M$_\odot$ pc$^{-2}$, given the lack of galaxies in this region of the parameter space, which is unavoidably affected by strong selection biases.

\subsection{Star-forming galaxies}

In Fig.\,\ref{fig2} and Fig.\,\ref{fig3}, we added the disks of star-forming galaxies. In Paper I, we found that star-forming galaxies define two separate sequences: one formed by star-dominated HSD disks (mostly Sa-Sc) and one by gas-dominated LSD disks (mostly Sd-dI). In the following, we first discuss the possible links between HSD disks and Es/S0s, then those between LSD disks and dEs/dSphs. Note that we use the terminology `LSD disks' because the dominant mass component of these galaxies (atomic gas) lies in a thin rotating disk, but their subdominant stellar component may potentially be a thicker disk and/or a more spheroidal structure.

\subsubsection{\textcolor{black}{Structural relations of high-mass galaxies}}

HSD disks are star-dominated ($M_{\rm gas}/M_{\rm bar} \lesssim 0.3)$, so their positions in the structural planes do not vary substantially when considering stellar or baryonic quantities. \textcolor{black}{This is evident in Fig.~\ref{fig_corner_HSD}, which shows that the mass-size relations of stellar and baryonic HSD disks are consistent with each other within less than 1$\sigma$.}
In the mass-size planes, HSD disks lie towards the left of the Es/bulge sequence, having systematically smaller masses and mean surface densities. \textcolor{black}{Indeed, Fig.~\ref{fig_corner_HSD} shows that the mass-size relations of HSD disks and Es/bulges are statistically different at about 3$\sigma$.} Clearly, HSD disks could `move' towards the sequence of Es/bulges only through major mergers \citep[e.g.][]{Toomre1977}, increasing their baryonic mass and mean surface density, as well as altering their internal kinematic structure. We note, however, that the majority of Es probably form and quench in the early Universe ($z\simeq10$) through very rapid processes \citep[e.g.][]{McGaugh2024}, whereas HSD disks may form at a slower pace \citep[e.g.][]{Schombert2019-SFH}, so the two types of galaxy may simply form and evolve through different channels.

HSD disks largely overlap with S0 disks, indicating that the former may naturally evolve into the latter as they run out of gas. \textcolor{black}{Indeed,  Fig.~\ref{fig_corner_HSD} shows that the mass-size relations of S0 disks and HSD baryonic disks are consistent with each other within about 2$\sigma$. The agreement becomes even more significant (about 1$\sigma$) when we consider the HSD stellar disks, neglecting the gas component.} This result is in agreement with previous photometric studies \citep[e.g.][]{Kormendy2012} and dynamical ones \citep[][]{Rizzo2018-S0, Shelest2020}. It also supports the classification scheme proposed by \citet{vandenBergh1976-classification}, in which S0s and spirals form two parallel sequences.

\subsubsection{\textcolor{black}{Structural relations of low-mass galaxies}}

LSD disks are gas-dominated ($M_{\rm gas}/M_{\rm bar} \simeq 0.3-1.0)$, so their positions in the structural planes can substantially vary when considering stellar or baryonic quantities. In the baryonic planes (Fig.\ref{fig2}), LSD disks lie to the right of the dEs/dSphs/UDGs sequence, having slightly higher masses and mean surface densities. In the stellar mass-size plane (Fig.\,\ref{fig3}), instead, there is a clear overlap between LSD galaxies and dEs/dSphs/UDGs in the mass range $M_\star\simeq10^{7}-10^{9}$ M$_\odot$. \textcolor{black}{This is confirmed by Fig.~\ref{fig_corner_LSD}, which shows that the mass-size relation of dEs/dSphs/UDGs is different from that of baryonic LSD disks at about $3\sigma$, but virtually identical to that of stellar LSD disks.} Notably, several LSD galaxies in the SPARC sample may be classified as `gas-rich UDGs' but they are actually `classical' Sd-to-dI galaxies, indicating again that the UDG definition is arbitrary.

\textcolor{black}{One may wonder whether our results might be affected by projection effects because dEs/dSphs/UDGs in galaxy clusters are possibly thicker than star-forming dwarfs \citep[e.g. ][]{Chen2023-thickness, SanchezJanssen2019-shape, Rong2020-UDGshape}. However, we find that dEs/dSphs/UDGs with different apparent axis ratios follow the same relation with merely small differences in their observed scatter, eliminating the possibility of projection effects.}

Gas-dominated LSD galaxies with $M_\star\simeq10^{5}-10^{7}$ M$_\odot$ are not well represented in the SPARC sample but are known to exist \citep[see, e.g. Figure 1 in][]{Lelli2022}, so it is likely that the overlap between LSD galaxies and dEs/dSphs/UDGs persists at lower masses. This is in line with the known fact that the stellar component of star-forming dwarfs is similar to that of passive dwarfs \citep[e.g.][]{Bingelli1994, Ferguson1994}.

At the other end of the mass spectrum, LSD disks with $M_\star\simeq10^{10}$ M$_\odot$ do not have a clear `passive counterpart' with similar structural properties. These massive LSD galaxies are known to live in low-density environments \citep[e.g.][]{Bothun1993-LSB_environment, Bothun1997-LSBG}, so they may just remain star-forming galaxies across the whole Hubble time. 
These massive LSD galaxies should not be confused with the so-called `giant low-surface-brightness' (GLSB) galaxies, whose prototype is Malin 1 \citep{Bothun1987}. GLSB galaxies are rare systems with a double structure: an inner high-surface-brightness component (such as a star-dominated HSD disk and/or a stellar bulge) and an outer extreme LSD disk \citep{Lelli2010, Hagen2016, Saburova2019, Saburova2021}. GLSB galaxies are not included in our sample.

\section{Discussion}\label{sec_diss}

In the previous section, we present the structural planes between $M_{\rm bar}$, $R_{\rm 50, bar}$ and $\Sigma_{\rm 50, bar}$ for both passive and star-forming galaxies, spanning over $\sim$8 orders of magnitude in mass. 

For star-forming HSD galaxies (Sa-Sc) and passive galaxies (Es, S0s, dEs/dSphs, UDGs), the baryonic and stellar planes are nearly the same because these objects are star-dominated. Indeed, we confirm a series of known results: (1) HSD disks overlap with S0 disks, so spiral galaxies may simply transform into lenticulars as they run out of gas; (2) there is a dichotomy between Es/bulges and dEs/dSphs; and (3) UDGs are an extension of the sequence of typical dEs/dSphs. Notably, extreme dwarf galaxies with $M_{\rm bar}\simeq10^5-10^7$ M$_\odot$, $R_{\rm 50, bar}>1$ kpc and $\Sigma_{\rm bar, 50}<1$ M$_\odot$ pc$^{-2}$ may be under-represented in current samples, given the potential selection effects in luminosity and surface brightness in this region of the parameter space.

For star-forming LSD galaxies (mostly Sd-dI), the structural relations change significantly if one considers baryonic or stellar quantities because these galaxies are gas dominated. Star-forming LSD galaxies are offset from passive dwarfs in the baryonic planes, but the two populations overlap when considering the stellar planes, especially for $M_\star\lesssim10^9$ M$_\odot$. The obvious interpretation is that passive dwarfs arise from star-forming dwarfs that have lost their gas component, moving across the baryonic planes but staying almost in the same position of the stellar planes. Importantly, differently from the possible evolution between spirals and lenticulars, the evolution from dIs to dEs/dSphs cannot be just a matter of converting gas into stars because the baryonic mass would be conserved. The cold gas needs to be physically removed from the galaxy, so the key question is what physical processes (internal or external) may be responsible for such gas removal.

Star-forming dwarfs consume their gas in a very inefficient way and have gas depletion times much longer than the Hubble time \citep{vanZee2001b, McGaugh2017}. In addition, there is overwhelming evidence that stellar feedback (stellar winds and supernovae) is unable to eject substantial amounts of gas out of the potential wells of dwarf galaxies \citep{Lelli2014b, Lelli2014a, Concas2017, Concas2019, Concas2022, McQuinn2019, Marasco2023}, so feedback-driven quenching is not a viable possibility. Star-forming dwarfs evolving in isolation will probably remain on the LSD sequence of the structural planes for most of their lifetime. The only way to remove the dominant gas component of star-forming dwarfs is to have some external mechanisms due to the environment. For example, there is mounting evidence for gas stripping in galaxy clusters due to a variety of different mechanisms, such as ram-pressure stripping, tidal stripping, and galaxy harassment \citep[e.g.][]{Boselli2014-cluster, Boselli2022}. Similar mechanisms may also be effective in galaxy groups in proximity of the central galaxy, as suggested by the strong morphology-density relation of dwarf galaxies in the Local Group \citep[e.g.][]{Mateo1998, Tolstoy2009}.

A long-standing issue with the possible evolution from dIs to dEs/dSphs is that the overlap in the stellar planes mostly occurs in the mass range where dEs/dSphs are nucleated, whereas dIs rarely show nuclei \citep{Cote2006, Lisker2007}. The simplest possibility is that star-forming dwarfs evolve into non-nucleated dwarfs, while nucleated dwarfs form a different category of galaxies \citep[e.g.][]{Jerjen1997}. Another possibility is that some star-forming dwarfs experience a starburst before losing their gas, creating a central nuclei and evolving into dEs(N) \citep{Lelli2014b}. The latter scenario is in agreement with the fact that starburst dwarfs have a more concentrated mass distribution (gas, stars, and dark matter) than typical dIs \citep{vanZee2001a, Lelli2012a, Lelli2012b}, but similar to those of the brigthest dEs \citep{Lelli2014b, Rys2014-Virgo}.

\section{Summary}\label{sec_conclusion}

We studied the structural relations between $M_{\rm bar}$, $R_{\rm 50, bar}$ and $\Sigma_{\rm 50, bar}$ for both passive and star-forming galaxies. The difference between baryonic and stellar structural relations is virtually negligible for star-dominated HSD galaxies (mostly Sa-Sc) and passive galaxies (Es, S0s, dEs/dSphs, UDGs), but it is very significant for gas-dominated LSD galaxies (mostly Sd-dI). Our results can be summarised as follows:
\begin{enumerate}
    \item \textcolor{black}{Passive stellar components form four distinct sequences or `families', namely (i) Es and bulges, (ii) S0 disks, (iii) non-nucleated dwarfs (dEs/dSphs/UDGs), and (iv) nucleated dEs.}
    \item Star-forming HSD disks overlap with S0 disks in both the baryonic and stellar planes, so spirals may simply turn into lenticulars as they convert all their available gas into stars.
    \item Star-forming LSD disks form a sequence that is offset from the one of dEs/dSphs/UDGs towards higher $M_{\rm bar}$ and $\Sigma_{\rm 50, bar}$, but the two galaxy populations overlap in the stellar planes, suggesting that star-forming dwarfs could evolve into passive dwarfs after gas removal due to some external mechanism.
    \item UDGs seem to be an extension of the sequence of non-nucleated dEs/dSphs; their progenitors may be typical star-forming LSD disks that lost their gas, with no need of a major expansion of their stellar component.
\end{enumerate}

In the coming years, the combination of wide-field \hi\ surveys from SKA pathfinders \citep[e.g. WALLABY,][]{Murugeshan2024} with near-infrared (NIR) photometric surveys will allow us to study the baryonic structural relations of galaxies for much larger galaxy samples and in a variety of different cosmic environments. A first step in this direction is represented by the up-coming BIG-SPARC database \citep{Haubner2024}, which will provide \hi\ and NIR data for several thousands of galaxies.

\begin{acknowledgements}
      The authors thank Konstantin Haubner and Illaria Ruffa for constructive discussions during this study. EDT was supported by the European Research Council (ERC) under grant agreement no. 101040751. 
\end{acknowledgements}

\bibliographystyle{aa}
\bibliography{MRrelation}

\begin{appendix}
\section{The best-fit parameters}\label{app_1}
\begin{figure*}
    \centering
    \includegraphics[width=0.85\linewidth]{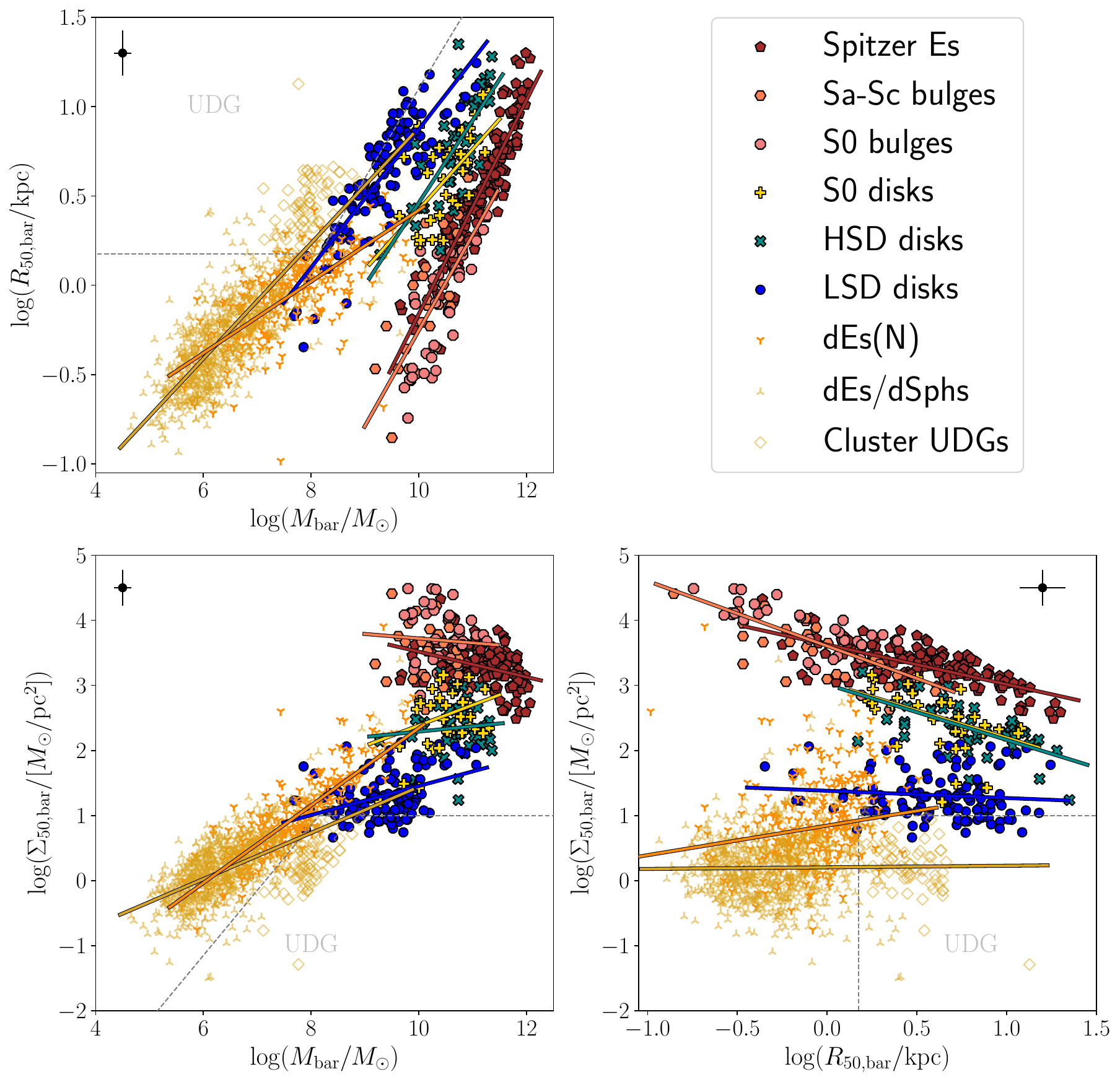}
    \caption{\textcolor{black}{Same as Fig.\,\ref{fig2} but we added the best-fit lines to the scaling relations of Es, bulges, S0 disks, HSD baryonic disks, LSD baryonic disks, dEs/dSphs/UDGs, and dEs(N), respectively. The best-fit line of each group is matched with the same colour.}}
    \label{fig_A1}
\end{figure*}

\begin{figure*}
    \centering
    \includegraphics[width=0.85\linewidth]{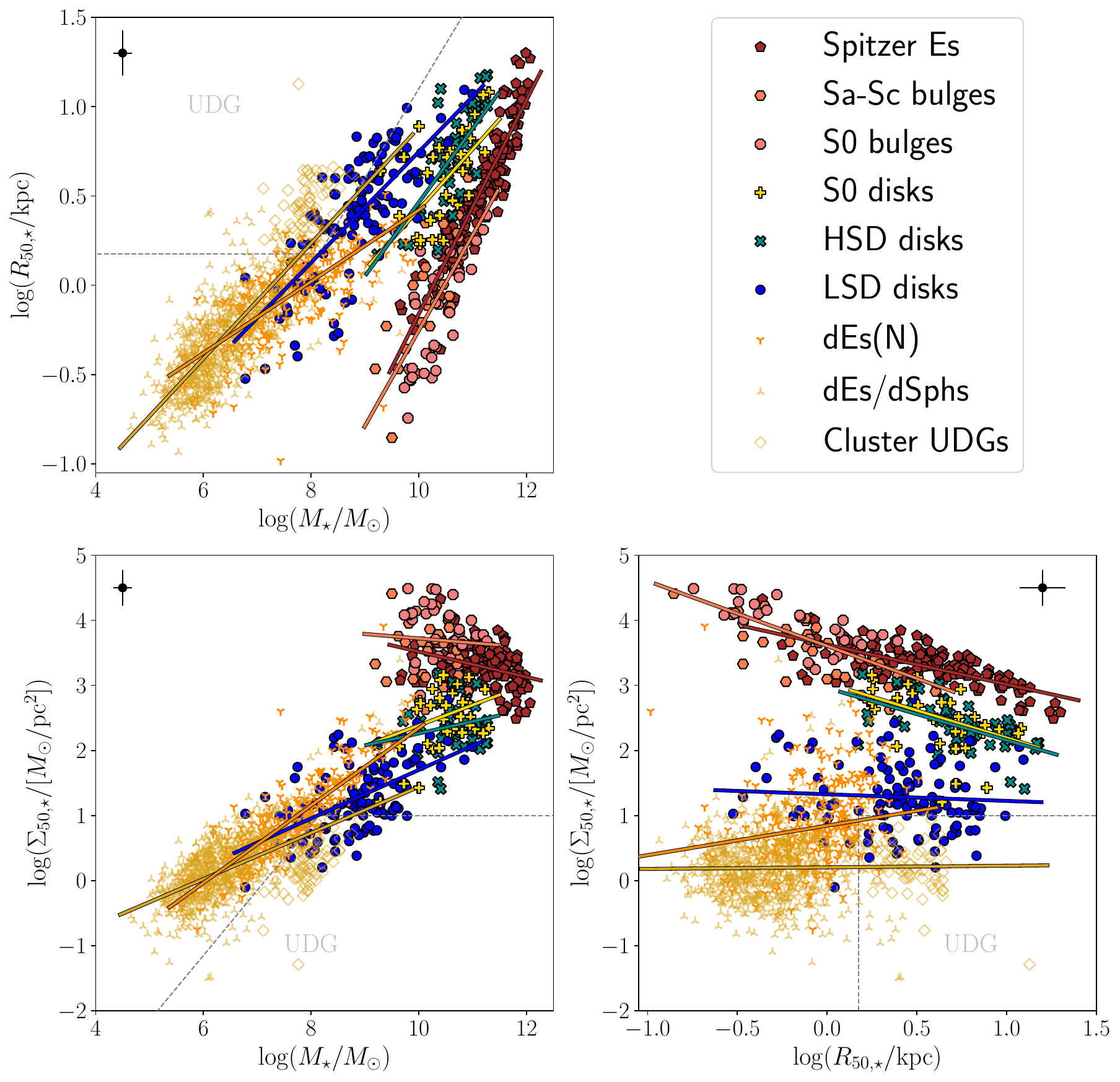}
    \caption{\textcolor{black}{Same as Fig.\,\ref{fig3} but added the best-fit lines to the scaling relations of Es, bulges, S0 disks, HSD stellar disks, LSD stellar disks, dEs/dSphs/UDGs, and dEs(N), respectively. The best-fit line of each group is matched with the same colour.}}
    \label{fig_A2}
\end{figure*}

We fitted the scaling relations of different galaxy populations adopting a power-law relation,
\begin{equation}\label{eq_fit}
    \log y = \alpha \cdot \log x + \beta,
\end{equation}
where $x$ and $y$ represent $M_{\rm \star}$, $R_{\rm 50, \star}$, or $\Sigma_{\rm 50, \star}$ , respectively. Following Paper I, we determined the best-fit $\alpha$ and $\beta$ using \textsc{roxy} \citep{Bartlett2023-roxy}. The results of different galaxy populations are shown in Table~\ref{tab_star_scaling}. We exhibit the best-fit lines \textcolor{black}{of the main galaxy populations in Figs.~\ref{fig_A1} and \ref{fig_A2}}. Given that we compiled structural parameters from several different studies, the errors on these quantities may not be entirely self-consistent. Thus, we prefer to perform the linear fits without considering the errors on the structural parameters. This choice implies that the intrinsic scatter provided by the \textsc{roxy} software is \textcolor{black}{basically the same as the observed scatter ($\sigma_{\rm obs}$), as shown in Table~\ref{tab_star_scaling}}.

\begin{table*}[h]
\begin{center}
    \caption{\textcolor{black}{\textsc{Roxy} best-fit parameters of the structural relations of different galaxy populations.}}
    \label{tab_star_scaling}
\begin{tabular}{c |c c c| c c c| c c c}
\hline
\hline
& \multicolumn{3}{c|}{$M_{\rm bar}-R_{\rm 50, bar}$ } 
& \multicolumn{3}{c|}{$M_{\rm bar}-\Sigma_{\rm 50, bar}$} 
& \multicolumn{3}{c}{$R_{\rm 50, bar}-\Sigma_{\rm 50, bar}$}\\
&
$\alpha$ & $\beta$ & $\sigma_{\rm obs}$ &
$\alpha$ & $\beta$ & $\sigma_{\rm obs}$ &
$\alpha$ & $\beta$ & $\sigma_{\rm obs}$ \\
\hline
Es &
$0.60^{+0.02}_{-0.02}$ &
$-6.15^{+0.26}_{-0.25}$ &
$0.14^{+0.01}_{-0.01}$ &
$-0.19^{+0.04}_{-0.05}$ &
$5.47^{+0.54}_{-0.46}$ &
$0.29^{+0.02}_{-0.02}$ &
$-0.61^{+0.05}_{-0.05}$ &
$3.63^{+0.04}_{-0.04}$ &
$0.22^{+0.01}_{-0.01}$ \\

bulges &
$0.53^{+0.06}_{-0.06}$ &
$-5.60^{+0.58}_{-0.63}$ &
$0.23^{+0.02}_{-0.02}$ &
$-0.07^{+0.13}_{-0.12}$ &
$4.41^{+1.25}_{-1.36}$ &
$0.45^{+0.04}_{-0.04}$ &
$-1.00^{+0.12}_{-0.12}$ &
$3.61^{+0.04}_{-0.04}$ &
$0.31^{+0.03}_{-0.03}$ \\

Es + bulges &
$0.64^{+0.02}_{-0.02}$ &
$-6.67^{+0.20}_{-0.21}$ &
$0.18^{+0.01}_{-0.01}$ &
$-0.28^{+0.03}_{-0.04}$ &
$6.51^{+0.41}_{-0.38}$ &
$0.36^{+0.02}_{-0.02}$ &
$-0.68^{+0.04}_{-0.04}$ &
$3.66^{+0.02}_{-0.02}$ &
$0.26^{+0.01}_{-0.01}$ \\

S0 disks &
$0.34^{+0.08}_{-0.08}$ &
$-2.95^{+0.82}_{-0.88}$ &
$0.20^{+0.03}_{-0.02}$ &
$0.31^{+0.17}_{-0.16}$ &
$-0.75^{+1.65}_{-1.77}$ &
$0.41^{+0.06}_{-0.05}$ &
$-0.84^{+0.29}_{-0.30}$ &
$3.04^{+0.18}_{-0.18}$ &
$0.38^{+0.05}_{-0.05}$ \\

dEs/dSphs &
$0.26^{+0.01}_{-0.01}$ &
$-1.96^{+0.05}_{-0.06}$ &
$0.17^{+0.00}_{-0.00}$ &
$0.48^{+0.02}_{-0.02}$ &
$-2.88^{+0.11}_{-0.11}$ &
$0.34^{+0.01}_{-0.01}$ &
$0.05^{+0.07}_{-0.07}$ &
$0.21^{+0.03}_{-0.03}$ &
$0.48^{+0.01}_{-0.01}$ \\

UDGs &
$0.18^{+0.06}_{-0.05}$ &
$-1.01^{+0.44}_{-0.45}$ &
$0.16^{+0.02}_{-0.01}$ &
$0.63^{+0.11}_{-0.11}$ &
$-4.74^{+0.86}_{-0.87}$ &
$0.31^{+0.03}_{-0.02}$ &
$-1.18^{+0.25}_{-0.25}$ &
$0.76^{+0.12}_{-0.12}$ &
$0.33^{+0.03}_{-0.03}$ \\

dEs/dSphs/UDGs &
$0.32^{+0.01}_{-0.01}$ &
$-2.35^{+0.05}_{-0.05}$ &
$0.19^{+0.00}_{-0.00}$ &
$0.35^{+0.02}_{-0.02}$ &
$-2.10^{+0.12}_{-0.11}$ &
$0.38^{+0.01}_{-0.01}$ &
$0.02^{+0.05}_{-0.05}$ &
$0.20^{+0.02}_{-0.02}$ &
$0.48^{+0.01}_{-0.01}$ \\

dEs(N) & 
$0.20^{+0.01}_{-0.01}$ & 
$-1.58^{+0.10}_{-0.10}$ & 
$0.16^{+0.01}_{-0.01}$ & 
$0.60^{+0.03}_{-0.03}$ & 
$-3.61^{+0.19}_{-0.21}$ & 
$0.32^{+0.01}_{-0.01}$ & 
$0.45^{+0.16}_{-0.16}$ & 
$0.84^{+0.04}_{-0.04}$ & 
$0.56^{+0.03}_{-0.02}$ \\

Baryonic HSD disks &
$0.46^{+0.06}_{-0.06}$ &
$-4.15^{+0.66}_{-0.65}$ &
$0.20^{+0.02}_{-0.02}$ &
$0.08^{+0.12}_{-0.13}$ &
$1.46^{+1.40}_{-1.29}$ &
$0.39^{+0.05}_{-0.04}$ &
$-0.85^{+0.16}_{-0.14}$ &
$3.01^{+0.12}_{-0.13}$ &
$0.31^{+0.03}_{-0.03}$ \\

Stellar HSD disks &
$0.41^{+0.06}_{-0.06}$ &
$-3.67^{+0.63}_{-0.64}$ &
$0.19^{+0.02}_{-0.02}$ &
$0.18^{+0.12}_{-0.13}$ &
$0.42^{+1.36}_{-1.22}$ &
$0.38^{+0.04}_{-0.04}$ &
$-0.80^{+0.16}_{-0.17}$ &
$2.96^{+0.14}_{-0.12}$ &
$0.33^{+0.04}_{-0.03}$ \\

Baryonic LSD disks &
$0.39^{+0.02}_{-0.02}$ &
$-3.00^{+0.20}_{-0.21}$ &
$0.16^{+0.01}_{-0.01}$ &
$0.22^{+0.04}_{-0.04}$ &
$-0.78^{+0.39}_{-0.41}$ &
$0.32^{+0.02}_{-0.02}$ &
$-0.11^{+0.11}_{-0.11}$ &
$1.38^{+0.08}_{-0.08}$ &
$0.36^{+0.03}_{-0.02}$ \\

Stellar LSD disks &
$0.31^{+0.03}_{-0.02}$ &
$-2.35^{+0.21}_{-0.23}$ &
$0.22^{+0.02}_{-0.01}$ &
$0.37^{+0.05}_{-0.05}$ &
$-2.03^{+0.45}_{-0.47}$ &
$0.44^{+0.03}_{-0.03}$ &
$-0.10^{+0.16}_{-0.14}$ &
$1.33^{+0.08}_{-0.08}$ &
$0.54^{+0.04}_{-0.04}$ \\
\hline
\end{tabular}
\tablefoot{For passive galaxies, the stellar scaling relations are virtually equivalent to the baryonic ones.}
\end{center}
\end{table*}

\end{appendix}

\end{document}